\documentclass[12pt,letterpaper,a4paper]{article}
\usepackage[includeheadfoot,
            marginratio={1:1,2:3},
            width=412pt,
            height=688pt,]{geometry}
\usepackage{amsmath}
\usepackage{amsfonts}
\usepackage{amssymb}
\usepackage{graphicx}
\usepackage{empheq}
\usepackage{color}
\usepackage{hyperref}

\usepackage{float}
\restylefloat{table}
\restylefloat{figure}

\newcommand{\nc}{\newcommand}
\nc{\beq}{\begin{equation}}
\nc{\eeq}{\end{equation}}
\nc{\bea}{\begin{eqnarray}}
\nc{\eea}{\end{eqnarray}}
\newcommand{\eq}[1]{\begin{equation}
                     \begin{split} #1 \end{split}
                     \end{equation}}
\def\ov{\overline}
\def\cO{{\cal O}}

\def\cF{\mathcal{F}}

\def\cO{\mathcal{O}}

\newdimen\csize\csize=1.5ex
\def\young#1{\tiny\vcenter{\hbox{\vrule\vtop{\hrule
  \offinterlineskip\halign{&\vbox
  {\hbox to\csize {\strut\hss##\hss\vrule}\hrule}\cr#1 \crcr}}}}}

           \usepackage{amsmath}
\usepackage{amsfonts}
\usepackage{amssymb}
\usepackage{stmaryrd}
\usepackage{graphicx}
\usepackage{empheq}
\usepackage{paralist}

\usepackage{graphicx}
\usepackage{multirow}
\usepackage{color}

\nc{\lb}{\llbracket}
\nc{\rb}{\rrbracket}
\nc{\gl}{\llbracket}
\nc{\gr}{\rrbracket}

\allowdisplaybreaks[2]
\numberwithin{equation}{section}

\begin{document}

\vspace*{-1.5cm}
%\begin{flushright}
% {\small
%  MPP-2014-....\\
%  }
%\end{flushright}

\vspace{1.5cm}
\begin{center}
{\LARGE
Combining Universal and Odd RR Axions for Aligned Natural Inflation}
\vspace{0.4cm}

\end{center}

\vspace{0.35cm}
\begin{center}
 Xin Gao$^{\dag,\ddag}$\footnote{Email: xingao@vt.edu}, Tianjun Li$^{\dag, \flat}$\footnote{Email: tli@itp.ac.cn}  and Pramod Shukla$^\sharp$\footnote{Email: pkshukla@to.infn.it}
\end{center}

\vspace{0.1cm}
\begin{center}
{\it
%$^{1}$ Max-Planck-Institut f\"ur Physik (Werner-Heisenberg-Institut), \\
%   F\"ohringer Ring 6,  80805 M\"unchen, Germany  \\
%\vspace{0.4cm}

$^{\dag}$ State Key Laboratory of Theoretical Physics
and Kavli Institute for Theoretical Physics China (KITPC),
      Institute of Theoretical Physics, Chinese Academy of Sciences,
Beijing 100190, P. R. China \\
\vspace{0.4cm}
$^{\ddag}$ Department of Physics, Robeson Hall, 0435,  Virginia Tech,  \\
850 West Campus Drive, Blacksburg, VA 24061, USA\\
\vspace{0.4cm}
$^{\flat}$ School of Physical Electronics,
University of Electronic Science and Technology of China,
Chengdu 610054, P. R. China \\
\vspace{0.4cm}
%$^2$ Consortium for Fundamental Physics, Physics Department, Lancaster University, LA1
%4YB, United Kingdom\\
%\vspace{0.4cm}
$^{\sharp}$ Universit\'a di Torino, Dipartimento di Fisica and I.N.F.N. - sezione di Torino \\
Via P. Giuria 1, I-10125 Torino, Italy
}

\vspace{0.2cm}

\vspace{0.5cm}
\end{center}

\vspace{1cm}

%%%%%%%%%%%%%%%%%%%%%%%%%%%%%%%%%%%%%%%%%%%%%%%
%%%%%%%%%%%%%%%%%%%%%%%%%%%%%%%%%%%%%%%%%%%%%%%
%%%%%%%%%%%%%%%%%%%%%%%%%%%%%%%%%%%%%%%%%%%%%%%
%%%%%%%%%%%%%%%%%%%%%%%%%%%%%%%%%%%%%%%%%%%%%%%
%%%%%%%%%%%%%%%%%%%%%%%%%%%%%%%%%%%%%%%%%%%%%%%
%%%%%%%%%%%%%%%%%%%%%%%%%%%%%%%%%%%%%%%%%%%%%%%
%%%%%%%%%%%%%%%%%%%%%%%%%%%%%%%%%%%%%%%%%%%%%%%
%%%%%%%%%%%%%%%%%%%%%%%%%%%%%%%%%%%%%%%%%%%%%%%

\begin{abstract}

We successfully embed the  Kim-Nilles-Peloso (KNP) alignment mechanism for enhancing the axion decay constant
in the context of large volume type IIB orientifolds. The flat direction is generated in the plane of $(C_0-C_2)$
axions corresponding to the involutively even universal axion $C_0$ and odd axion $C_2$, respectively.
The moduli stabilization with large volume scheme has been established as well. %An extension of this KNP-type potential with a combination of quadratic monodromy potential have been embedded for realizing significantly large running of spectral index needed for reconciling Planck and BICEP2 data.

\end{abstract}

\clearpage

%\tableofcontents

%%%%%%%%%%%%%%%%%%%%%%%%%%%%%%%%%%%%%%%%%%%%%%%
%%%%%%%%%%%%%%%%%%%%%%%%%%%%%%%%%%%%%%%%%%%%%%%
%%%%%%%%%%%%%%%%%%%%%%%%%%%%%%%%%%%%%%%%%%%%%%%
%%%%%%%%%%%%%%%%%%%%%%%%%%%%%%%%%%%%%%%%%%%%%%%
%%%%%%%%%%%%%%%%%%%%%%%%%%%%%%%%%%%%%%%%%%%%%%%
%%%%%%%%%%%%%%%%%%%%%%%%%%%%%%%%%%%%%%%%%%%%%%%
%%%%%%%%%%%%%%%%%%%%%%%%%%%%%%%%%%%%%%%%%%%%%%%
%%%%%%%%%%%%%%%%%%%%%%%%%%%%%%%%%%%%%%%%%%%%%%%

%\newpage

\section{Introduction and Motivation}
\label{sec_intro}
The recent BICEP2 results \cite{Ade:2014xna} have undoubtably shaken the status of inflationary model building in string cosmology. The discovery of primordial B-mode polarization of the cosmic microwave background has been recently claimed by the BICEP2 Collaboration. This claim can be understood as a signature of gravitational wave which is encoded in the so-called tensor-to-scalar ratio ($r$). The BICEP2 observations fix the inflationary scale by ensuring a large tensor-to-scalar ratio $r$
as follows~\cite{Ade:2014xna}
\bea
\label{eq:cosmo-I}
& &  \hskip0.5cm r = 0.20^{+0.07}_{-0.05} ~(68\% ~{\rm CL}) ~,~\nonumber\\
& & H_{\rm inf} \simeq 1.2 \times 10^{14} \, \left(\frac{r}{0.16}\right)^{1/2} \, \, {\rm GeV} \, ,
\eea
where $H_{\rm inf}$ denotes the Hubble parameter during the inflation. Subtracting the various dust models and re-deriving the $r$ constraint still result in high significance of detection and one has $ r=0.16^{+0.06}_{-0.05} $.
In order to reconcile the tension between  BICEP2 \cite{Ade:2014xna} result and
PLANCK \cite{Ade:2013zuv}, WMAP data \cite{Hinshaw:2012aka}, it demands the following windows for the other cosmological observables
\bea
\label{eq:cosmo-II}
& & \hskip-1cm  \ln\left(10^{10} \, \, P_s \right) = 3.089^{+0.024}_{-0.027}, \, \, \, \, n_s = 0.957 \pm 0.015 , \, \, \, \, \alpha_{n_s} = - 0.022^{+0.020}_{-0.021} ~,~
\eea
where $P_s$ is the scalar power spectrum and $\alpha_{n_s}$ is the running of spectral index $n_s$. All these cosmological observables can be written out in terms of the inflationary potential and its various derivatives. Thus, with the available experimental data from various sources so far, the shape of a single field inflationary potential is significantly constrained. As a reverse computation, writing out various derivatives of inflationary potential in terms of the aforementioned cosmological observables, a generic single field inflationary potential can be locally
reconstructed \cite{Choudhury:2014kma, Ma:2014vua, Gao:2014pca}.

In order to  realize the required large value of tensor-to-scalar ratio $r$, the inflaton field needs to travel over trans-Planckian distance according to the famous Lyth bound \cite{Lyth:1996im}. Further, it also suggests the inflationary process to be (a high scale process) near the scale of Grand Unified Theory (GUT). As a result, embedding inflationary models in a UV complete framework, such as string theory, is inevitable.  The UV sensitivity in chaotic inflation class of models has been recently addressed in \cite{Kaloper:2014zba}. The BICEP2 claims, if proven, can also provide invaluable pieces of information in search of a consistent supersymmetry (SUSY) breaking scale \cite{Ibanez:2014zsa, Harigaya:2014pqa}.

With large field excursions, the other relevant issues from higher order corrections should also be taken care of for the viability of the model
 \cite{Chialva:2014rla, Baumann:2011nm, Burgess:2014tja}. If the BICEP2 result is confirmed, it would serve as a huge discriminator filtering out many among the plethora of inflationary models developed so far. However, it is interesting that the three
 classes of inflationary models, namely the chaotic-type
 \cite{Linde:1983gd, Kawasaki:2000yn, Kallosh:2010ug, Nakayama:2013txa, Kobayashi:2014jga, Gao:2014fha},
 natural-type \cite{Freese:1990rb, ArkaniHamed:2003wu, Kim:2004rp, Czerny:2014wza} as well as
 Assisted/N/M-flation type \cite{Liddle:1998jc, Copeland:1999cs, Jokinen:2004bp, Dimopoulos:2005ac, Grimm:2007hs, Ashoorioon:2009wa, Ashoorioon:2011ki, Cicoli:2014sva} are among the winners. In the context of models developed in a purely string framework prior to the BICEP2 results, the
 axion monodromy inflation \cite{McAllister:2008hb, Flauger:2009ab} was found to be much closer (but still insufficient)
 %but still insufficient as per
 to fulfill the BICEP2 claims.
 There have been very vibrant and speedy progress on these lines of developing chaotic- or (multi)natural- type of  inflationary models utilizing axion monodromy in a very short post-BICEP2 period so far \cite{Palti:2014kza, Blumenhagen:2014gta, Grimm:2014vva, Arends:2014qca, Hassler:2014mla, Czerny:2014qqa, Ellis:2014rxa, Hebecker:2014eua,  Marchesano:2014mla, Tye:2014tja, Ben-Dayan:2014zsa, Long:2014dta, Kappl:2014lra, Choi:2014rja, Li:2014owa, Li:2014xna, McAllister:2014mpa, Ashoorioon:2014jja, Kobayashi:2014ooa, Higaki:2014sja}.

Regarding the axionic inflation in Type IIB string framework,
LARGE Volume Scenario (LVS) \cite{Balasubramanian:2005zx}
%the model are preferred to be developed in the context of LARGE Volume Scenarios (LVS)\cite{Balasubramanian:2005zx}.
provides a well-controlled moduli stabilization mechanism, which makes
the lightest moduli  a good candidate for being an inflaton.  In the LVS mechanism,
%exhibiting an exponentially large volume for the internal Calabi-Yau threefold.
 the exponentially large volume of the internal Calabi-Yau threefold ${\cal V}$ is favored as it also provides a control over the (un-)known $\alpha^\prime$ \cite{Becker:2002nn} as well as string loop corrections $g_s$ \cite{Berg:2007wt, Cicoli:2007xp}. In fact, it has been observed that the known/conjectured forms of these corrections, at the level of K\"ahler potential, appear with volume suppressed terms in the scalar potential \cite{Balasubramanian:2005zx, Berg:2007wt, Cicoli:2007xp}, which makes large volume scenario more robust as well. Further due to the presence of (extended-)no-scale structure in the context of type IIB swiss-cheese compactification, various volume moduli directions orthogonal to the overall Calabi-Yau (CY) volume ${\cal V} $ remain flat. The breaking of flatness in the orthogonal direction via (non-)perturbative corrections leads to a flat enough inflationary potential; for example, see the models with inflaton being identified with divisor volume moduli \cite{Conlon:2005jm, Bond:2006nc, BlancoPillado:2009nw, Cicoli:2008gp, Cicoli:2011ct, Blumenhagen:2012ue, Lust:2013kt, Gao:2014fva}.

In the context of LVS framework, embedding of axion monodromy type potential has been recently proposed in \cite{Blumenhagen:2014gta}, in which the universal axion $C_0$ could drive the inflationary process.
Based on certain assumptions on the background flux, the large volume expansion has been argued to be useful for  trusting the effective field theory (EFT) description even in a non-perturbative regime where the string coupling $g_s$ satisfies $1< g_s <10$ \cite{Blumenhagen:2014gta}.
On the other hand, in the context of axionic inflationary models of natural-type inflation \cite{Freese:1990rb}, a large decay constant has been proposed to be realized in a Kim-Nilles-Peloso (KNP)-type two-field potential \cite{Kim:2004rp}.
The main idea is to align two sub-planckian decay constants such that with a certain rotation of field basis, one could create a hierarchy in the decay constant of the newly constructed axionic basis.
The best advantage of  this type of axionic inflation is that unlike N-flation \cite{Dimopoulos:2005ac, Grimm:2007hs, Cicoli:2014sva} which requires a large number of (${\cal O}(10^3-10^4)$) axions assisting the inflationary process, this is a two-field model.
However, the standard KNP-model with two fields usually requires large anomaly coefficients or equivalently large gauge groups of the non-perturbative effects to generate the potential, which is challenging to be embedded into a realistic particle physics or string model.
On these lines, the standard KNP-model has been generalized to N-fields (with $N<10$) \cite{Choi:2014rja,Higaki:2014pja} to facilitate the axionic alignments (as well as to keep the number of axions less than those required in N-flation model).

Motivated by the KNP proposal for enhancing the decay constant to trans-planck scale, in this article, we propose a new class of inflationary
potentials in the context of LVS framework. The inflationary direction lies in the plane of $(C_0 - C_2)$ axions, where $C_0$ corresponds to the involutively even universal axion while $C_2$ is involutively odd axion.
If we restrict the orientifold to be divisor exchange or reflection,
in order to support large volume scenarios with the orientifold odd axions, the underline Calabi-Yau threefold should have $h^{1,1}(CY_3) \geq 3$
\cite{Gao:2013pra, Gao:2013rra}. %\footnote{There is only one example for $h^{1,1}(CY_3)=2$ which can have orientifold odd cohomology. The odd moduli stabilization through F-term is studied in \cite{Gao:2013rra}.
%, where we take the universal axion stabilized at tree level.}.
Using two such involutively odd axions and magnetized non-perturbative effects, recently a KNP-type scenario has been proposed in \cite{Long:2014dta}. Unlike this proposal, we utilize the universal axion $C_0$ along with a single odd axion $C_2$ to get the required alignment for the natural inflation. This engineering solves one of the major challenges of \cite{Blumenhagen:2014gta} by taking the framework within perturbative regime as large enough decay constant is realized within $g_s <1$ in our model.
%Moreover, as the decay constant of $c_0$ axion is given in terms of string coupling $g_s$ as $f_{c_0} = \frac{g_s}{\sqrt 2}$ which is corrected by volume suppressed terms, it consistently fits in large volume scenario framework.
Moreover, a combination of $C_0$ and $C_2$ axions provides a better decoupling in the kinetic sector unlike the case with two odd axions \cite{Long:2014dta}.

%our model is different from \cite{Long:2014dta} in the sence that the inflation is driven by two different kinds of moduli,  one of which preserves a shift symmetry at tree level.

The article is organized as follows. In section \ref{sec_Setup} we provide a brief and relevant feature of type IIB orientifolds. Section \ref{sec_KNP}  summarizes the original KNP formalism \cite{Kim:2004rp} for enhancing the decay constant in a two-field potential. In section \ref{sec_AxionMonodromyV}, we  provide a successful embedding of KNP-type potential in large volume scenarios with the inclusion of odd axion along with universal axion. Section \ref{sec_numerical} presents a detailed numerical analysis with a couple of benchmark models. Finally, in section \ref{sec_conclusions} we provide a summary with possible open challenges.

\section{Relevant Ingredients of Type IIB Orientifolds}
\label{sec_Setup}

We consider type IIB superstring theory compactified on an orientifold of a
Calabi-Yau threefold $CY_3$ with $O3/O7-$plane. The full orientifold action is
$\cO= (-)^{F_L}\Omega_p \sigma$, where the $F_L$ is the spacetime fermion number in the left-moving sector,  $\Omega_p$ denotes world-sheet parity while $\sigma$ denotes a holomorphic and isometric involution.
By performing the detailed dimensional reduction from ten to
four dimensions \cite{Lust:2006zg}, the low energy effective action at the second order in derivatives is
given by a supergravity theory, whose dynamics is encoded
in three building blocks; namely the K\"{a}hler potential $K$, the holomorphic superpotential $W$, and the
holomorphic gauge kinetic functions. These building blocks can be generically written in terms of appropriate ${\cal N} =1$ coordinates ($S, G^a, T_\alpha$) defined as
 \eq{
\label{eq:N=1_coords}
S&=i \, c_0+ e^{-\phi} \, , \, \, \, \, {G}^a = i \,  c^a - S \, {b}^a \, , \nonumber}
\eq{
T_\alpha&=\frac{1}{2} \kappa_{\alpha\beta\gamma}\, t^\beta t^\gamma +  \,
  i \left(\rho_\alpha -\frac{1}{2}\kappa_{\alpha a b} \, {c^a b^b}\right)
-\frac{1}{4}\, e^\phi \,  \kappa_{\alpha ab}\, {\bar G}^a {(G+\bar G)}^b\, ,
}
where $t^\alpha$ is the two-cycle volume while $c_0$, $c^a$ and $\rho_\alpha$ correspond to RR axions $C_0$, $C_2$, and $C_4$, respectively. Further, $\kappa_{\alpha \beta \gamma}$ and $\kappa_{\alpha a b}$ are triple intersections numbers of the even/odd two cycle. Here, the indices $\alpha$ run in even (1,1)-cohomology of CY orientifold  ($H^{1,1}_+(CY_3/\sigma)$) while indices $a$ are counted in odd (1,1)-cohomology $H^{1,1}_-(CY_3/\sigma)$.

\subsubsection*{The K\"ahler Potential $K$}

Generically, the K\"{a}hler potential is given as
\eq{
\label{eq:K}
K = - \ln\left(S+{\bar S}\right)
-\ln\left(-i\int_{X}\Omega_3\wedge{\bar\Omega_3}\right)-2\ln\left( {\cal Y}\,
(S, G^a, T_\alpha,...)\right)~,~\,
}
where ${\cal Y}= \frac{1}{6}{\cal K}_{ABC}\, t^A t^B t^C$ is the volume
of the Calabi-Yau manifold expressed in terms of two-cycle
volumes $t^A$.
The dots in \eqref{eq:K}  denote the potential appearance of other moduli
%with correspoonding shift in dilaton fileds, e.g.
like D3/D7-brane fluctuations (and
hence complex structure moduli which get coupled after including
brane-fluctuations) or Wilson line moduli.
Unfortunately, ${\cal Y}$ is only implicitly given in terms
of the chiral superfields.  It is in general non-trivial
to invert the last relation in \eqref{eq:N=1_coords}, and so it is not possible to write $K$ in terms of $T_{\alpha}$ explicitly. Further, the most general K\"ahler potential can also depend on the derivatives of chiral superfield  \cite{Chialva:2014rla, Baumann:2011nm}. However, we ignore such higher order corrections in the present analysis.

\subsubsection*{The Superpotential $W$}
The general schematic form of the superpotential $W$ is given as
\bea
\label{eq:W}
W & &= \int_{X}G_3\wedge\Omega + \sum_{D} {\cal A}_{D}(z^{\tilde a}, G^a, \,{\cal F}_D\,, ...)
\, e^{-\, a_D^\alpha T_{\alpha}}  \nonumber\\
& &= W_{\rm cs} + W_{np}~,~\,
\eea
where the first term is the Gukov-Vafa-Witten (GVW)
three-form flux induced tree-level superpotential \cite{Gukov:1999ya} (See \cite{Dasgupta:1999ss,Taylor:1999ii} also for related work).
The second term denotes a sum over  non-pertur\-ba\-tive corrections coming from the
Euclidean $D3$-brane instantons or gaugino condensation on D7-branes
\cite{Witten:1996bn}. Again the dots indicate a further dependence on e.g.
D3/D7-brane fluctuations or Wilson line moduli. Further, the prefactor contains not only the one-loop
Pfaffian for fluctuations around the instanton background but also
contributions from so-called (gauge-)fluxed instantons \cite{Grimm:2011dj, Kerstan:2012cy}
and Euclidean $D1$-brane instantons \cite{Grimm:2007xm}. The presence of gauge fluxes on the divisor contributing to the non-perturbative superpotential helps in alleviating \cite{Grimm:2011dj} the chirality issue proposed in \cite{Blumenhagen:2007sm}. It also helps in stabilizing  all the odd moduli, with or without the help of poly-instanton effects~\footnote{The proper zero mode structure of poly-instanton in type IIB orientifold has been clarified in \cite{Blumenhagen:2012kz}. }
\cite{Gao:2013rra}. Also, in principle one has to sum over all the possible instanton or gaugino condensation effects, and in the presence of extra magnetic fluxes turned-on on the relevant odd two-cycles sitting inside the relevant divisor, this issue becomes more delicate in terms of satisfying tadpole/anomaly cancellation conditions, etc \cite{Bianchi:2011qh, Berglund:2012gr}; see also a related review in \cite{Blumenhagen:2009qh}. However in the present study, our approach would be rather phenomenological, and we consider the most suitable ansatz for the superpotential without getting into these technicalities.

We would consider the gauge flux effects on the orientifold invariant divisor $D$ (having involutively odd two-cycles and) contributing to the non-perturbative superpotential.
Usually, there are two kinds of non-perturbative corrections, one is induced through Euclidean D3-brane (E3-brane) instanton while the other one through gaugino condensation. For E3-brane instanton, we require the gauge flux turned-on on the brane to be $\cF_E \in H^{1,1}_{-}(D_E)$ in order to ensure the instanton to be of $O(1)$-type.
For gaugino condensation with a stack of $2N$ D7-branes, they should also be placed at orientifold invariant positions. If the D7-branes coincide with an O-plane, i.e. both $N$ branes and their images are placed on top of an $O7^{-/+}$-plane, it provides $SO(2N)/SP(2N)$ gauge group dynamics. If the $D7$-branes and their images  wrap on the same internal geometry, it yields $SP(2N)/SO(2N)$ gauge group.  Turning on a gauge flux with ${\cal F}_D \in H_-^{1,1}(D)$, the fluxed brane can remain invariant under the orientifold projection, while turning on gauge flux ${\cal F}_D \in H_+^{1,1}(D)$ breaks the gauge symmetry to a unitary group \cite{Blumenhagen:2012kz, Blumenhagen:2008zz}, then a D-term will be generated by the $U(1)$ subgroup with Fayet-Iliopoulos terms. Since $O7^{-/+}$-plane carries $-8/+8$ times of the D7-brane charge, in the following, we always assume that we have $O7^{-}$-plane and turn-on only the odd gauge flux $\cF_D \in H^{1,1}_{-}(D)$ on the branes. Also, for the time being, we concentrate on the F-term dynamics and just consider the suitable form of superpotential $W$ with multiple gaugino condensation configuration. The D-brane tadpole cancellation as well as the zero-modes condition are assumed to be settled when addressed in concrete setup, and one can start with the following form of the (odd moduli $G^a$ dependent) holomorphic prefactor ${\cal A}_{D}(z^{\tilde a}, G^a, \,{\cal F}_D\,, ...)$ of (\ref{eq:W}),
\bea
\label{eq:Wgen}
& & \hskip-2cm W_{np} = A\, \sum_{{\cal F}_D} \, e^{-\, a_D^\alpha\, T_{\alpha}} \, \exp\left[ - \, a_D^\alpha \, \,  h_1({{\cal F}_D})  \, \, S \, - \, a_D^\alpha\, h_2({{\cal F}_D}) \, G^a \right]~,~\,
\eea
where $h_i({{\cal F}_D})$s are gauge flux dependent constants turned-on along the odd two-cycles of the divisor $D$ supporting the non-perturbative superpotential contribution. This form of superpotential will be heavily utilized in the upcoming sections.

\subsubsection*{The Scalar Potential $V$}
From the  K\"{a}hler potential and the superpotential one can compute the
${\cal N}=1$ scalar potential
\bea
\label{eq:Vgen}
V = e^{K}\Biggl(\sum_{I,\,J}{K}^{I\bar{J}} {\cal D}_I W
{\bar{\cal D}}_{\bar J } {\bar W} - 3 |W|^2 \Biggr)~,~\,
\eea
where the sum  runs over all moduli. For studying the K\"{a}hler moduli dynamics, we will assume that
the complex structure moduli and dilaton have
already been stabilized supersymmetrically  as ${\cal D}_{c.s.} W = 0, \, {\cal D}_{S} W = 0$. Further, on the lines of \cite{Blumenhagen:2014gta}, we assume that with the freedom available through the landscape of background fluxes, one can still keep universal axion $c_0$ massless or `nearly' massless. We will quantify what we mean by `nearly' and elaborate on this point later while considering the explicit computations in sections \ref{sec_AxionMonodromyV} and \ref{sec_numerical}.

\section{Review of KNP-Type Natural Inflation}
\label{sec_KNP}
Let us very briefly review the original KNP proposal for natural inflation \cite{Kim:2004rp}.
We consider the following two-field inflationary potential
\bea
\label{eq:KNP}
& & V(\phi_1, \phi_2) = \sum_{i = 1}^2 \, \Lambda_i \,  \left(1 - \cos\left[\frac{\phi_1}{f_i} + \frac{\phi_2}{g_i} \right]\right) ,
\eea
where $f_i$ and $g_i$'s can be sub-Planckian decay constants as the most natural choice. The determinant of the Hessian of this potential is simplified to
\bea
& & {\rm Det}\left(V_{ij}\right) = \frac{(f_2 \, g_1 - f_1 \, g_2)^2 \, \prod_{i = 1}^{2} \, \Lambda_i \,  \cos\left[\frac{\phi_1}{f_i} + \frac{\phi_2}{g_i} \right] \, }{f_1^2 \, f_2^2 \, g_1^2 \, g_2^2}~.~\,
\eea
Thus, it will have a flat direction if the following condition holds
\bea
\label{eq:cond}
& & \frac{f_1}{f_2} = \frac{g_1}{g_2}~.~\,
\eea
Therefore, a small enough deviation from this condition can create a mass hierarchy between the two axions rotated in a new basis. As we will see explicitly in a moment, one can elegantly create a mass hierarchy and (with appropriate axionic rotation) an alignment leading to the enhancement of decay constant of the lighter combination also occurs. With
 the following rotation of axions
\bea
& & \psi_1 = \frac{g_1 \, \phi_1 + f_1 \, \phi_2}{\sqrt{f_1^2 + g_1^2}}\, , \, \, \, \psi_2 = \frac{f_1 \, \phi_1 - g_1 \, \phi_2}{\sqrt{f_1^2 + g_1^2}}~,~\,
\eea
we reformulate the expression eq.(\ref{eq:KNP}) as under
\bea
& & V(\psi_1, \psi_2) =   \Lambda_1 \,  \left(1 - \cos\left[\frac{\psi_1}{f_1^\prime} \right]\right) + \Lambda_2 \,  \left(1 - \cos\left[\frac{\psi_1}{f_2^\prime} + \frac{\psi_2}{f_{\rm eff}} \right]\right) \, ,
\eea
where $f_1^\prime, \, f_2^\prime$ and $f_{\rm eff}$ take the form as below
\bea
& & f_1^\prime = \frac{f_1 \, g_1}{\sqrt{f_1^2 + g_1^2}} \, , \, \, f_2^\prime = \frac{f_2 \, g_2 \, \sqrt{f_1^2 + g_1^2}}{f_1\, f_2 + g_1\, g_2} \, , \, \, \, f_{\rm eff} = \frac{f_2 \, g_2 \, \sqrt{f_1^2 + g_1^2}}{|f_1\, g_2 -  g_1\, f_2|}~.~\,
\eea
Thus, if the deviation from the flatness condition eq.(\ref{eq:cond}) is small enough, one can generate an `effectively' large decay constant for $\psi_2$ combination. Further, together with eq.(\ref{eq:cond}) and an appropriate hierarchy $\Lambda_2 \ll \Lambda_1$, one can make the field $\psi_1$ heavier than $\psi_2$ with the respective masses at the minimum given as
\bea
& & m_{\psi_1}^2 \simeq \Lambda_1 \left(\frac{1}{f_1^2} +\frac{1}{g_1^2} \right) , \, \, \, m_{\psi_2}^2 \simeq \frac{\Lambda_2 \, (f_2 \, g_1 - f_1 \, g_2)^2}{g_2^2 \, f_2^2 \, \left(f_1^2 + g_1^2\right)}~.~\,
\eea
Stabilizing $\psi_1$  at one of its minimum $ \ov \psi_1 = 0$ would result in a single axion potential
with large decay constant as below
 \bea
& & V(\psi_2) = \Lambda_2 \,  \left(1 - \cos\left[ \frac{\psi_2}{f_{\rm eff}} \right]\right) \,.
\eea
Now we turn to the embedding of KNP-type mechanism in large volume scenario. The main focus would be to utilize universal RR axion $C_0$ along with an involutively odd RR axion $C_2$.

\section{Realizing Natural Inflation in Large Volume Scenarios}
\label{sec_AxionMonodromyV}
Let us consider the following ansatz for the K\"ahler potential $K$ motivated by the large volume scenarios. After introducing a single odd modulus $G^1$ via the appropriate choice of orientifold involution\footnote{For constructing explicit examples of CY orientifold with $h^{11}_-(CY_3/{\cal O}) \ne 0$,
see \cite{Blumenhagen:2008zz, Gao:2013pra}.}, the K\"ahler potential becomes \cite{Gao:2013rra}
\bea
\label{eq:oddK}
& &\hskip-1cm K \equiv K_{cs} -\ln(S + \ov S)  - 2\, {\rm ln} \,{\cal Y} \\
& & \hskip0.5cm = K_{cs} -\ln(S + \ov S)  -2 \, {\rm ln} \, \left(\xi_B \, \Sigma_B^{3/2} -\xi_S \, \Sigma_S^{3/2} + {{\cal C}_{\alpha^\prime}} \right) ~,~\,\nonumber
\eea
where
\begin{eqnarray}
\label{eq:Yodd1}
 & & \Sigma_\alpha = T_\alpha + {\bar T}_\alpha + \frac{ \kappa_{\alpha11}}{2(S+ {\bar S})}({G^1}+{\bar G^1})({G^1}+{\bar G^1})~ \, \, {\rm for} \,\,\, \alpha\in\{B,S\},
 %& & \Sigma_S = T_S + {\bar T}_S + \frac{ \kappa_{S11}}{2(S+ {\bar S})}({G^1}+{\bar G^1})({G^1}+{\bar G^1}) ~,~\,\nonumber
\end{eqnarray}
and ${{\cal C}_{\alpha^\prime}}=- \frac{\chi(X)\xi(3)}{4 (2 \pi)^3 \, {g_s}^{3/2}}$.
This form of K\"ahler potential explicitly shows the shift symmetries in various RR axionic directions; namely the universal axion $C_0$, the involutively even axion $C_4$ and the involutively odd axion $C_2$. Although the presence of $\alpha^\prime$-corrections break the ``no-scale structure", it still leaves the direction orthogonal to ${\cal V}$ (which is $\tau_s$) to remain flat. This flatness and axionic shift symmetries are broken via the non-perturbative effects appearing in the following racetrack form of the superpotential which comes from eq.(\ref{eq:Wgen})
\bea
\label{eq:oddW}
& & \hskip-1cm W=  W_{\rm cs} \, +  A_0\,\, e^{-\, a_0 T_S}  \, \\
& & +  A_s\,\, e^{-\, a_s \left(T_S + h_1({\cal F}) \, S +\, h_2({\cal F}) \, G^1\right)}\,  - \, B_s \, \, e^{-\, b_s \left(T_S + h_3({\cal F}) \, S \, + h_4({\cal F}) \, G^1\right)}~,\nonumber
\eea
where
\bea
W_{cs} = W_{\rm cs1}  \, + \, S\, \, W_{\rm cs2} \, ~.~
\eea
Such a form of superpotential eq.(\ref{eq:oddW}) could be thought of arising from different stacks of unfluxed  and fluxed $D7$-branes wrapping the so-called small divisor in an orientifold invariant way as discussed before eq.(\ref{eq:Wgen}). As a result, we can set the gaugino condensations effects with $a_0=\frac{2 \pi}{N_0}, \, a_s=\frac{2 \pi}{N_1}$,  $b_s=\frac{2 \pi}{N_2}$, where $N_0$, $N_1$ and $N_2$ being the ranks of the corresponding gauge groups coming from different stacks of $D7$-branes. Further, $W_{cs1}, \, W_{cs2}, \, A_0, \, A_s$ and $B_s$  are generically complex structure moduli and background flux dependent quantities.  For the time being, these are considered to be constants as in the standard moduli stabilization schemes. At the outset, let us clearly mention the following inherent assumptions before coming to the scalar potential computation
\begin{itemize}
\item{In addition to background fluxes, there are gauge fluxes turned-on on the small divisor which induces axio-dilaton $S$ and odd axion $G^1$ dependence on top of the non-perturbative effects. These are encoded in such gauge flux dependent quantities $h_i({\cal F}) \, , \, \, \, \forall i \in \{1, 2, 3, 4\}$. For the minimal setting $h_1(\cF)$ and $h_3(\cF)$ are quadratic in gauge flux while $h_2(\cF)$ and $h_4(\cF)$ are linear in gauge flux. As a result, we should keep $h_1>h_2$ and $h_3>h_4$ %\footnote{In principle, one should sum over all the flux contributions. However, the qualitative results are similar \cite{Grimm:2011dj,Kerstan:2012cy,Gao:2013rra}.}
.}
\item{On the lines of \cite{Blumenhagen:2014gta}, we assume that in the absence of non-perturbative corrections to the superpotential, the landscape of background fluxes can facilitate one with keeping the universal RR axion $C_0$ massless (or  at least nearly massless) via creating a mass-hierarchy between dilaton mass and universal axion $C_0$.
Although the univesal axion appears as a linear term in the superpotential, by  tuning the background flux dependent parameters in the tree-level superpotential,  the $c_0$ axion shift symmetry remain (nearly) unbroken via the quadratic term induced in the scalar potential. In order to restore large volume scenarios as well as a decoupled KNP-type inflationary potential of $c_0-c^1$ axion, the coefficient $w_2$ in $W_{cs} = w_1 + c_0 \, w_2$ has to satisfy the following bound
\bea
|w_2|^2 \ll {\cal O} \left( \frac{e^{-a_s \, h_1/g_s}}{{\cal V}^3}\right) \sim {\cal O} \left( \frac{e^{-b_s \, h_3/g_s}}{{\cal V}^3}\right) ; \, \, \, \, \, w_1 \sim {\cal O}(1).
\eea
This is probably the strongest assumption in our model and should be examined to be realized in a concrete Calabi-Yau orientifold construction.}
\item{Based on the aforementioned point, we assume the standard procedure to stabilize the complex structure moduli and dilation via the background flux superpotential. So we naively consider $W_{cs} = w_1 + \, c_0 \, w_2$ such that $w_2 \ll w_1$ and we will quantify how small $w_2$ should be to trustfully recover the large volume potential.}
\end{itemize}
Utilizing these pieces of information, the F-term scalar potential can be computed from eq.(\ref{eq:Vgen}) for the given ansatz of $K$ and $W$, and various terms can be categorically collected as follows
\bea
& & \hskip-3cm V({\cal V}, \tau_s; \rho_s, b^1, c^1, c_0) \simeq V_{\rm LVS}({\cal V}, \tau_s; \, \rho_s, b^1) \\
& & \hskip2cm + V_{\rm rest}({\cal V}, \tau_s \, \rho_s, b^1; c^1, c_0) ~,~\,\nonumber
\eea
where $V_{\rm LVS}({\cal V}, \tau_s; \, \rho_s, b^1)$ is the large volume potential contributing at the leading order ${\cal O}\left(\frac{1}{{\cal V}^3}\right)$. With stabilizing the axions at one of their minima $\rho_s = 0 = b^1$ \cite{Flauger:2009ab, Long:2014dta}\footnote{This is also similar to the case of moduli stabilization via fluxed-instanton superpotential leading to the appearance of theta function in $W_{np}$. In that case, there are several extrema in the axionic directions due to theta-function periodicities appearing in the potential \cite{Gao:2013rra}, and $b^1=0$ is the simplest local minimum. }, the stabilized values of volume moduli is obtained by the solutions of following coupled expressions
\bea
\label{eq:constraint}
& & C_{\alpha^\prime} \simeq \frac{32 \, \sqrt2 \, a_0 \, \xi_S \, \ov\tau_s^{5/2} (a_0 \ov\tau_s -1)}{\left(1-4 a_0 \, \ov\tau_s\right)^2} ; \nonumber\\
& & \ov{\cal V} \simeq -\frac{6 \sqrt2 \, \xi_S \, W_{cs} \, \sqrt{\ov\tau_s} \, (a_0 \, \ov\tau_s-1) }{a_0 \, A_0 \, (4 a_0 \, \ov\tau_s -1)} \, e^{a_0 \, \ov\tau_s} \, \, ~.
\eea
Let us mention an important point that in our approach of stepwise moduli stabilization, one has to maintain the hierarchy $|V_{\rm LVS}| \gg |V_{\rm rest}|$ throughout and so one has to be careful while samplings of model dependent parameters are made. As we will see later, relatively larger gauge groups $N_1$ and $N_2$ are needed for realizing large decay constant, and in order to stabilize the overall volume of the CY to order $\cO(10^{3})$, we need $N_0 < N_{1,2}$ and then to maintain the mass hierarchy between standard K\"ahler moduli and universal axion together with odd moduli, one has to appropriately choose the gauge flux parameters $h_1$ and $h_3$ large enough.

\begin{figure}[H]
\begin{center}
\hspace*{-1.1cm} \includegraphics[width=14.5cm]{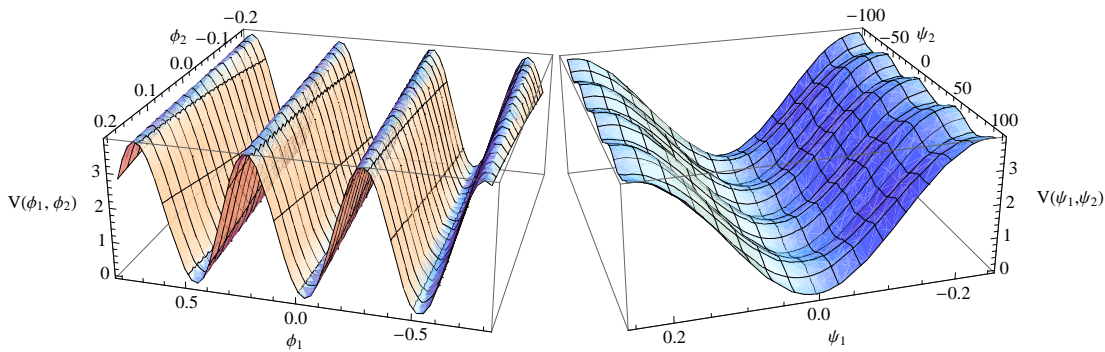}
\caption{ The two field potentials $V(\phi_1, \phi_2)$  and $V(\psi_1, \psi_2)$  (multiplied by $10^8$) respectively given in eq.(\ref{eq:Vaxion2}) and eq.(\ref{eq:Vaxion3}) are plotted for the sampling eq.(\ref{eq:samp1}). The second figure shows the enhanced decay constant for $\psi_2$ direction as compared to the sub-Planckian ones shown in the first figure.}
\label{Fig1}
\end{center}
\end{figure}

After stabilizing the heavier moduli and orientifold even axion $C_4$ along with odd axion $B_2$, the potential reduces to the form as below
\bea
\label{eq:Vexpand}
& & \hskip-0.1cm V_{\rm rest}({\cal V}, \tau_s \, \rho_s, b^1; c^1, c_0) \equiv V_{\rm rest}(c^1, c_0) \nonumber \\
& & \hskip1cm \simeq \Delta_0 + \Delta_1 \cos\left[ a_s \, h_1 \, c_0 + a_s \, h_2 \, c^1\right]  \nonumber \\
& & \hskip2cm + \Delta_2 \cos\left[ b_s \, h_3 \, c_0 + b_s \, h_4 \, c^1\right] + ...... \, ,
\eea
Thus, at the sub-leading order, the shift symmetry for the odd axion $c^1$ is broken. Here in the aforementioned simplification, the coefficients $\Delta_1$ and $\Delta_2$ are suppressed by factors $e^{- a_s \, h_1/g_s}$ and $e^{- b_s \, h_3/g_s}$ respectively as compared to $|V_{\rm LVS}|$ while $\Delta_0$ is the collection of all the terms independent of $c^1$ and $c_0$ axions.  Given our assumption that the coefficient of quadratic potential for universal axion $c_0$ generated at tree level can be fairly negligible by utilizing the flux freedom, and so $V_{\rm rest}$ is effectively the leading contribution to break the $c_0$ flatness.  Further, note that the model dependent parameters $h_1 $ and $h_3$ depend on the gauge flux ${\cal F}$ supported on the divisor with odd two-cycles contributing to the non-perturbative superpotential. Further, the dots denote those terms which are doubly suppressed by flux dependent exponentials and hence are subleading for small string coupling regime. Now after using an appropriate uplifting mechanism, one can rearrange the terms to result in the desired KNP-type potential \cite{Kim:2004rp}
\bea
\label{eq:Vaxion2}
& & \hskip-2cm V(\phi_1, \, \phi_2) \simeq  \Lambda_1 \left(1 - \cos\left[ \frac{n_1 \, \phi_1}{f_1} + \frac{n_2 \, \phi_2}{f_2} \right]  \right) \\
& & \hskip2cm +  \Lambda_2 \left(1 - \cos\left[ \frac{m_1 \, \phi_1}{f_1} + \frac{m_2 \, \phi_2}{f_2} \right]  \right) ~,~\,
\nonumber
\eea
where $\Lambda_i$'s can be collected in terms of model dependent parameters given as below
\bea
\label{eq:lambda12}
& & \Lambda_1 \simeq \frac{\sqrt2 \, a_0 \, a_s \, |A_0| \, |A_s| \, \ov\tau_s}{\, \xi_S \, \ov{\cal V} \, (a_0 \, \ov\tau_s -1)} \, {\rm Exp}\left[{- a_0 \ov\tau_s - a_s \ov \tau_s - \frac{a_s \, h_1}{g_s} }\right] \nonumber\\
& & \hskip3cm \simeq \frac{12 \, |W_{cs}| \, a_s \, |A_s| \, \ov \tau_s}{\ov{\cal V}^2 \, \xi_S \, (4 \, a_s \, \ov \tau_s -1 )} \, e^{- a_s \ov \tau_s - \frac{a_s \, h_1}{g_s} } \, ~,~\\
& & \hskip0cm \, \, \, \Lambda_2 \simeq \frac{\sqrt2 \, a_0 \, b_s \, |A_0| \, |B_s| \, \ov\tau_s}{\, \xi_S \, \ov{\cal V} \, (a_0 \, \ov\tau_s -1)} \, {\rm Exp}\left[{- a_0 \ov\tau_s - b_s \ov \tau_s - \frac{b_s \, h_3}{g_s} }\right]\nonumber\\
& & \hskip3cm \simeq \frac{12 \, |W_{cs}| \, b_s \, |B_s| \, \ov \tau_s}{\ov{\cal V}^2 \, \xi_S \, (4 \, b_s \, \ov \tau_s -1 )} \, e^{- b_s \ov \tau_s - \frac{b_s \, h_3}{g_s} } \, ~.~\nonumber
\eea
Further, in expression eq.(\ref{eq:Vaxion2}) of the potential, $n_i = a_s \, h_i $ and $m_i = b_s  \, \, h_{i+2}$ for $i = 1,2$. Subsequently, the canonically normalized fields $\phi_1$ and
$\phi_2$ are defined as follows
\bea
\label{eq:canonical}
& & \phi_1 \equiv c_0 \, f_{1} \simeq c_0 \, \frac{g_s}{\sqrt2}, \, \, \, \phi_2 \equiv  c^1 \, f_{2} \simeq c^1 \frac{\sqrt{- 3\, g_s \, \kappa_{B11}} \, \, {\xi_B}^{1/3}}{{\cal V}^{1/3}} ~.~\,
\eea
With the following redefinitions of the two-fields (similar to the original KNP-formalism reviewed in the last section)
\bea
\label{eq:rotat}
& & \psi_1 = \frac{n_1 \, f_2 \, \phi_1 + n_2 \, f_1 \, \phi_2}{\sqrt{n_1^2 \, f_2^2 +\, n_2^2 \, f_1^2}} , \, \,  \psi_2 = \frac{n_2 \, f_1 \, \phi_1 - n_1 \, f_2 \, \phi_2}{\sqrt{n_1^2 \, f_2^2 +\, n_2^2 \, f_1^2}}~,~\,
\eea
the expression of axionic potential eq.(\ref{eq:Vaxion2}) can be adjusted into the form as below
\bea
\label{eq:Vaxion3}
V(\psi_1, \psi_2) = \Lambda_1 \left(1 - \cos\left[ \frac{\psi_1}{f_1^\prime}\right]  \right) + \Lambda_2 \left(1 - \cos\left[ \frac{\psi_1}{f_2^\prime}+\frac{\psi_2}{f_{\rm eff}}\right]  \right)~,~\,
\eea
where
\bea
\label{eq:feff}
& & f_1^\prime = \frac{f_1 \, f_2}{\sqrt{n_1^2 \, f_2^2 +\, n_2^2 \, f_1^2}} ~,~\, f_2^\prime = \frac{f_1 \, f_2 \, \sqrt{n_1^2 \, f_2^2 + n_2^2 \, f_1^2}}{{\left(n_1\, m_1 \, f_2^2 +\, n_2 \, m_2 \, f_1^2\right)}} ~,~\, \nonumber
\eea
and
\bea
& & f_{\rm eff} = \frac{\sqrt{n_1^2 \, f_2^2 + n_2^2 \, f_1^2}}{|n_1 \, m_2 - n_2 \, m_1|}~.~\
\eea
Assuming a reasonable hierarchy $\Lambda_2 \ll \Lambda_1$, a justified diagonalization follows with a heavy ($\psi_1$) and a light ($\psi_2$) axionic combination. Further, stabilizing the heavier axion at its minimum $\ov \psi_1 = 0$ leads to a single-field natural inflation driven by a trans-Planckian axion as below
\bea
\label{eq:Vaxion4}
V(\psi_2) = \Lambda_2 \left(1 - \cos\left[\frac{\psi_2}{f_{\rm eff}}\right]  \right)~.~\,
\eea
Before  the detailed numerical analysis towards inflationary aspects, let us exemplify the moduli stabilization part by providing a benchmark sampling as below
\bea
\label{eq:samp1}
& & W_{cs} = -12, \, N_0 = 15, \, \xi_B = 1 = \xi_S, C_{\alpha^\prime} = 4.6, \, A_0 = 0.1, g_s = 0.35, \kappa_{B11} = -1,\nonumber\\
& & N_1 =30, N_2 = 32, h_1 = 15, h_2 =1, h_3 = 16, h_4 = 1, A_s =10, B_s =1.
\eea
Using these samplings in eq.(\ref{eq:constraint}), eq.(\ref{eq:lambda12}), eq.(\ref{eq:canonical}) and eq.(\ref{eq:feff}), one gets
\bea
\label{eq:samp2}
 & & \ov{\cal V} \simeq 925.7, \, \, \, \ov\tau_s \simeq 2.99 \, , \, \, \, f_1 \simeq 0.248, \, f_2 \simeq 0.105 ,\, \nonumber\\
& & |V_{\rm LVS}| \simeq 5.0 \times 10^{-7} \, , \, \, \Lambda_1 \simeq 1.78 \times 10^{-8}, \, \Lambda_2 \simeq 1.73 \times 10^{-9}\,,\, \, \nonumber\\
& & f_1^\prime \simeq 0.078, \, f_2^\prime \simeq 0.078, \, \, f_{\rm eff} \simeq 8.131.
\eea
Here, the parameters are chosen such that one gets
\begin{itemize}
\item{The effective decay constant $f_{\rm eff} > 7$. It is the minimal values to fit the  BICEP2 data. To match the PLANCK result, we can relax the constrain to  $f_{\rm eff} > 4$ as we will systematically explore in the numerical section. For the sampling eq.(\ref{eq:samp1}) and eq.(\ref{eq:samp2}), the enhancement of decay constant can be seen from Fig.\ref{Fig1}.}
\item{$\Lambda_2 \simeq 10^{-9}$, which is needed to have a high inflationary Hubble scale $H_{\rm inf} \simeq 10^{14}$GeV as per the requirement of constraints given in eq.(\ref{eq:cosmo-I}) or equivalently the magnitude of scalar power spectrum being $P_s \simeq 2.2 \times 10^{-9}$ given in eq.(\ref{eq:cosmo-II}).}
\item{Two hierarchies: $|V_{\rm LVS}| \gg V_{\rm rest}$ and $\Lambda_1 \gg \Lambda_2$. These are implicitly needed for reaching the single field inflationary potential via step-by-step route with hierarchial check-points.}
\end{itemize}

\subsection*{On validity of the single field inflationary approach}
\begin{figure}[H]
\centering
\includegraphics[scale=0.7]{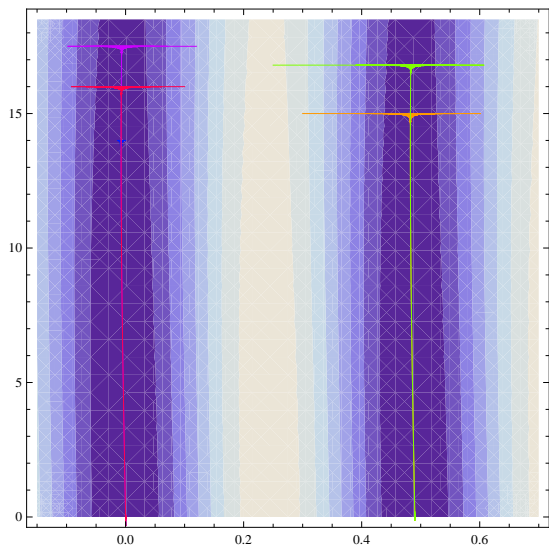}
\caption{The various inflationary trajectories for five initial conditions. The two minima in dark blue are separated by maxima in light blue shade.}
\label{trajectory}
\end{figure}

\begin{figure}[H]
\centering
 \includegraphics[width=0.47\textwidth]{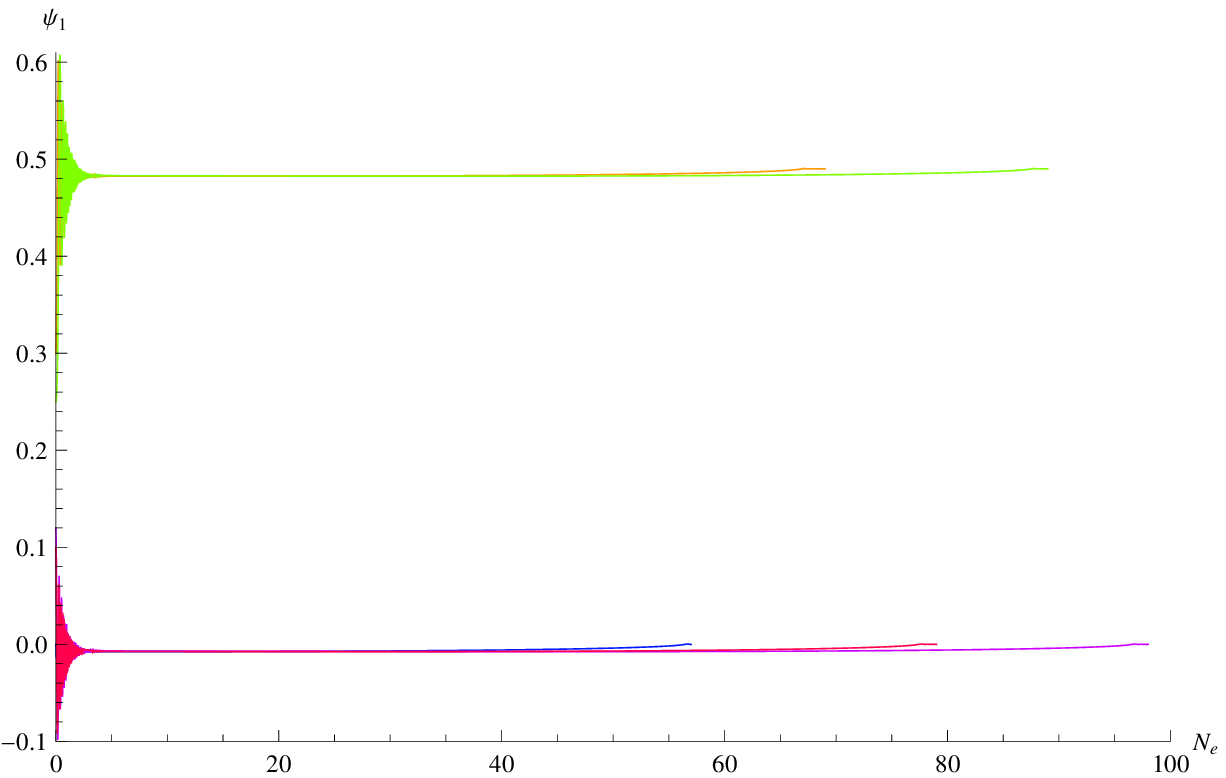}
 \includegraphics[width=0.45\textwidth]{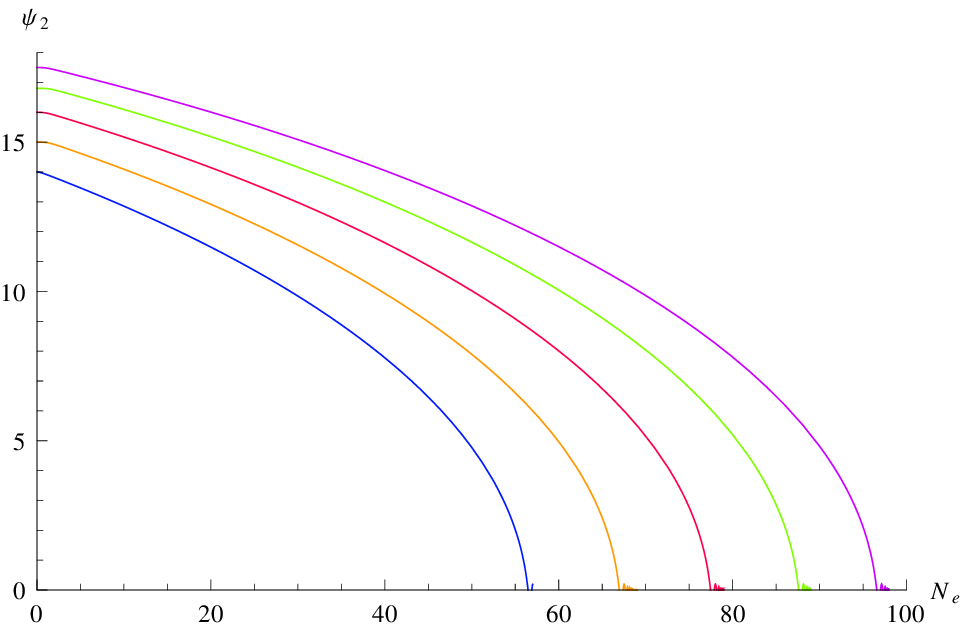}
% \\
% \includegraphics[width=0.4\textwidth]{oscillate3.eps}
% \includegraphics[width=0.4\textwidth]{oscillate4.eps}
\caption{Evolution of heavier ($\psi_1$) and lighter ($\psi_2$) axionic combinations during inflationary process reflecting its single field nature..}
\label{psiS}
\end{figure}

\begin{figure}[H]
\centering
\includegraphics[scale=0.8]{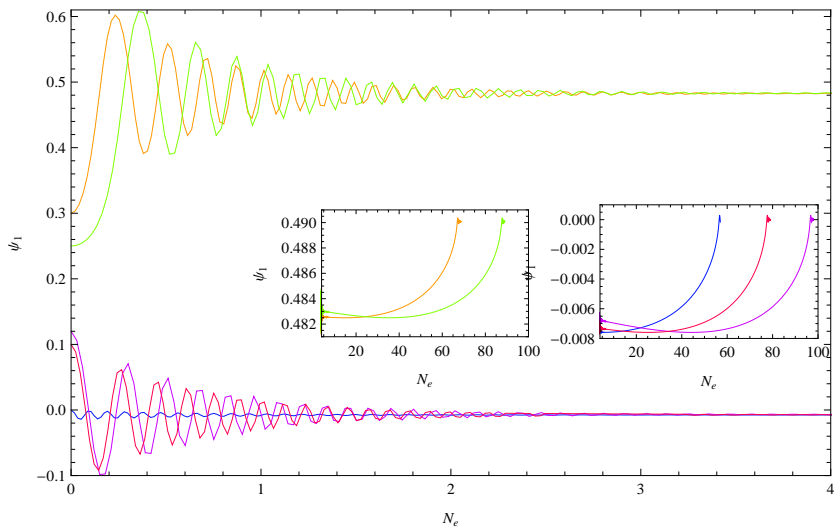}
\caption{A closer look at the evolution of heavier field $\psi_1$ showing a negligible shift from their respective minima in each of the five inflationary trajectories. The shift in heavier field $\psi_1$ is less than 0.01 within the inflationary regime.}
\label{psi1detailed}
\end{figure}

To justify that single field approach is a valid description, i.e. the heavier fields, set to their respective minima, do not significantly get shifted while the lightest axionic combination is in inflating phase, let us consider the dynamics of next-to-light field $\psi_1$ and perform the field evolution analysis of two-field inflationary potential (\ref{eq:Vaxion3}). The dynamics is governed by the Einstein-Friedmann equations given as
\begin{subequations}
\bea
\label{Friedmann}
\frac{d^2}{dN^2}\psi^a+{\Gamma^a}_{bc}\frac{d\psi^b}{dN}\frac{d\psi^c}{dN}+\left(3+\frac{1}{H}\frac{dH}{dN}\right)\frac{d\psi^a}{dN}+\frac{{\cal G}^{ab} \partial_b V}{H^2}=0 ,
\eea
\bea
\label{constran}
H^2=\frac{1}{3}\left(V(\psi^a)+\frac{1}{2}H^2 \, {\cal G}_{ab} \frac{d\psi^a}{dN}\frac{d\psi^b}{dN} \right) ,
\eea
\end{subequations}
where we use the background $N$ e-folding number as independent evolution coordinate with $dN = H dt$. Here, let us recall that given the shift symmetry in $C_0$ and $C_2$ directions in the K\"ahler potential, for the canonically normalized fields $\phi^a$s (with $a = 1, 2$), one has the kinetic matrix ${\cal G}^{ab}={\delta}^{ab}$. It holds in the new $\psi^a$ basis as well \footnote{A very quick check for kinetic sector metric to be diagonal with ${\cal G}_{ab}={\delta}_{ab}$ in $\psi^a$ basis is: $ds^2 = d\phi_1^2 + d\phi_2^2 = d\psi_1^2 + d\psi_2^2 = {\delta}_{ab} d\psi^a d\psi^b$ using relations in (\ref{eq:rotat}).}. So all Christophel connections vanish. In addition, using expressions (\ref{Friedmann}) and (\ref{constran}), one can derive another useful expression for variation of Hubble rate in terms of e-folding,
\bea
\label{third}
\frac{1}{H}\frac{dH}{dN}=\frac{V}{H^2}-3 .
\eea
Now let us numerically solve these evolution equations for model dependent sampling  given in (\ref{eq:samp1} - \ref{eq:samp2}), and see the dynamics of heavier ($\psi_1$) and lighter ($\psi_2$) axionic combinations during inflationary process. To get the full trajectories from a second order differential equations, we choose $\frac{d\psi^a}{dt}\frac{d\psi_a}{dt}|_{t=0}=0$ along with following five initial conditions for $\psi^a(0) \,, \, \, \forall \, \,  a \in \{1,2\}$,
\begin{table}[H]
  \centering
  \begin{tabular}{|c||c|c|c|c|c|}
  \hline
   & $I_1$& $I_2$ & $I_3$ & $I_4$ & $I_5$ \\
   \hline
   \hline
   $\psi_1(0)$  & 0.00 & 0.30 & 0.10 &  0.25 & 0.12 \\
   \hline
   $\psi_2(0)$  & 14.0 & 15.0 & 16.0 &  16.8 & 17.5 \\
   \hline
   $N_F$     & 57   & 69   & 79   &  89   & 98 \\
      \hline
  \end{tabular}
  \caption{Five initial conditions studied for sampling given in (\ref{eq:samp1} - \ref{eq:samp2}).}
  \label{initial}
 \end{table}
Utilizing the numerical solutions, various inflationary trajectories are plotted in Figure \ref{trajectory}. This figure also shows that depending on the initial condition, the field $\psi_1$ gets settled in its nearest minimum, for example trajectory with initial condition $\psi_1 =0.1$ settles in the minimum at zero while the one with $\psi_1 =0.3$ settles in a nearby minimum. Moreover, Figure \ref{trajectory} also confirms that heavier axionic combination does not get involved in the inflationary process, although it undergoes a negligible shift from its minimum. Except the first set of initial conditions $I_1$ which already corresponds to a single field inflation, for other trajectories $I_2-I_5$ one observes some initial oscillations in $\psi_1$ direction which is just an artifact of initial conditions $\frac{d\psi^a}{dt}\frac{d\psi_a}{dt}|_{t=0}=0$. These oscillations die off within a couple of e-foldings as seen in Figures [\ref{psiS}-\ref{psi1detailed}] and result in an effectively single field inflationary process.

This analysis has been done to justify a mass-hierarchy $m_{\psi_1} \gg m_{\psi_2}$ being maintained in such a way that the dynamics of two fields effectively lead to a single field inflationary process. The same has been done via designing a hierarchy $\Lambda_1 \gg \Lambda_2$ through appropriate sampling of model dependent parameters. %Based on similar logic, the hierarchy $|V_{\rm LVS}| \gg |V_{\rm rest}|$, which also means that $|V_{\rm LVS}| \gg \Lambda_1 \gg \Lambda_2$, should even better respect the mass hierarchy among the inflaton $\psi_2$ and the other set of heavier fields, throughout the inflationary process. Thus, this analysis indirectly supports our step-by-step approach to reach the single field Natural inflation potential (\ref{eq:Vaxion4}).

\section{Detailed Inflationary Investigations}
\label{sec_numerical}
\subsection*{Revisiting the Standard Natural Inflation}
Let us recall the relevant features of standard natural inflation by checking the consistency requirements of cosmological observables from the PLANCK and BICEP2 data. Usually it is qualitatively mentioned that the decay constant for axion utilized in the natural inflation must be trans-Planckian, i.e., much larger than the reduced Planck scale $M_{\rm Pl}$. The realization of large decay constant in string models has always been a challenge, and in one way or the other, the choice of model dependent parameters are crucially affected (and in confrontation within) to accommodate the observables in best possible manner. Therefore, we revisit this aspect to quantify the decay constant window needed to fulfill the minimal experimental requirements. For a given single field potential $V(\phi)$, the sufficient condition for ensuring the slow-roll inflation is encoded in a set of so-called slow-roll parameters defined as below
\bea
& & \epsilon \equiv \frac{1}{2} \, \left(\frac{V^\prime}{V}\right)^2 \ll 1 \, , \, \, \eta \equiv \frac{V^{\prime \prime}}{V} \ll 1 \, , \, \,  \xi \equiv \frac{V^\prime \, V^{\prime \prime\prime}}{V^2} \ll 1 \, ,
\eea
where $\prime$ denotes the derivative of the potential w.r.t. the inflaton field $\phi$. Also, the above expressions are defined in the units of reduced Planck mass $M_{\rm Pl}$ with $M_{\rm Pl} = 2.44 \times 10^{18} \, {\rm GeV}$.

The various cosmological observables such as the number of e-foldings $N_e$, scalar power spectrum $P_s$,
tensorial power spectrum $P_t$, tensor-to-scalar ratio $r$, scalar spectral index $n_s$,
and running of spectral index $\alpha_{n_s}$  can be written as the various derivatives of
the inflationary potential via introducing the aforementioned slow-roll parameters as follows
\bea
\label{eq:cosmo-III}
& & N_e \equiv \int_{\phi_{end}}^{\phi_*} \, \frac{1}{\sqrt{2 \epsilon}} \, d \phi \, , \nonumber\\
& & P_s \equiv   \left[\frac{H^2}{4 \, \pi^2 \, (2 \, \epsilon)} \, \left(1 - \left(2 \, C_E - \frac{1}{6}\right) \, \epsilon+\left(C_E \, -\frac{1}{3} \right) \, \eta \right)^2\right]  \, ,\nonumber\\
%& & P_t \equiv  8 \left[\frac{H^2}{4 \, \pi^2} \, \left(1 - (1 + C_E) \, \epsilon \right)\right] \, , \\
& & r  \simeq 16 \, \epsilon \left[1 -\frac{4}{3} \, \epsilon +\frac{2}{3} \, \eta + 2 \, C_E \, (2 \, \epsilon -\eta)\right] \, , \nonumber\\
& & n_s \equiv \frac{d \ln P_s}{d \ln k} \simeq 1 + 2 \biggl[\eta - 3 \, \epsilon -\left(\frac{5}{3} + 12 \, C_E \right)\, \epsilon^2 + (8 \, C_E -1) \epsilon \, \eta \nonumber\\
& & \hskip3.5cm + \frac{1}{3}\,\eta^2 -\left(C_E - \frac{1}{3} \right) \, \xi \biggr] \, ,\nonumber\\
& & \alpha_{n_s}\equiv \frac{d n_s}{d \ln k} \simeq 16 \, \epsilon \, \eta - 24 \, \epsilon^2 - 2 \, \xi \, , \nonumber
%& & n_t \equiv \frac{d \ln P_t}{d \ln k} \simeq  \, ,\nonumber\\
%& & n_r \equiv \frac{d r}{d \ln k} \simeq n_t - (n_s-1)  \, ,
\eea
where $C_E = -2 + 2 \ln 2 +\gamma \simeq -0.73$, $\gamma =0.57721$ being the Euler-Mascheroni constant.

For the standard singe field natural-inflation potential
\bea
\label{eq:Vaxion5}
V(\phi) = \Lambda_0 \left(1- \cos\left[\frac{\phi}{f}\right] \right),
\eea
the slow-roll parameters as well as the three main cosmological observables ($n_s, \, r $ and $\alpha_{n_s}$) to be constrained as per the relations in eq.(\ref{eq:cosmo-I}-\ref{eq:cosmo-II}) are simplified as below
\bea
& & \epsilon(\phi) = \frac{\cot\left[\frac{\phi}{2 \, f}\right]^2}{2 \, f^2}\, , \, \, \, \eta(\phi) = \frac{\cos\left[\frac{\phi}{f}\right] \,  \csc\left[\frac{\phi}{2 \, f}\right]}{2 \, f^2} \, , \, \, \, \xi(\phi)=- \frac{\cot\left[\frac{\phi}{2 \, f}\right]^2}{f^4} ~,~\, \\
& & \hskip-0.3cm  N_e(\phi) = - 2 \, f^2 \, \ln\left[\cos\left[\frac{\phi}{2 \, f}\right]\right] - N_e^{\rm end}~,~\,\nonumber\\
& &  r(\phi) = \frac{4 \, \left(-2 + 6\, C_E \, + 3\, f^2 - 3 \, f^2 \cos\left[\frac{\phi}{f}\right]\right) \, \cot\left[\frac{\phi}{2\, f}\right]^2 \,  \csc\left[\frac{\phi}{2 \, f}\right]^2 }{3 \, f^4} ~,~\,\nonumber\\
& & \hskip-0.2cm n_s(\phi) = - \biggl[\biggl\{17 + 60 \, C_E + 30 \, f^2 - 18 \, f^4 + 8\, \left(4 + 6 \, C_E - 3 \, f^2 + 3 \, f^4\right) \cos\left[\frac{\phi}{f}\right] \biggr\} \nonumber\\
& & \hskip2cm - \left(- 7 + 12 C_E + 6 f^2 + 6 f^4\right) \cos\left[\frac{2 \, \phi}{f}\right] \biggr]  \times \frac{\csc\left[\frac{\phi}{2 \, f}\right]^4}{48 \, f^4} \,  ~,~\,\nonumber\\
& &  \hskip-0.4cm \alpha_{n_s}(\phi) = -\frac{\csc\left[\frac{\phi}{2 \, f}\right]^6 \, \, \sin\left[\frac{\phi}{f}\right]^2}{2 \, f^4}~,~\,\nonumber
\eea
where $N_e^{\rm end} = f^2 \, \ln\left[1 - \frac{1}{2 \, f^2}\right]$ is evaluated at $\epsilon = 1$ where inflation ends.  Further, $N_e^{\rm end}$ values fall in the range $\{0.41, 0.50\}$ for decay constant lying inside $\{1, 16\}$.
\subsubsection*{Number of e-foldings $N_e$}
\begin{figure}[H]
\begin{center}
\includegraphics[width=6.8cm]{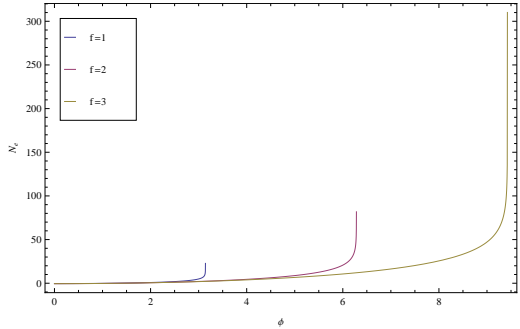}
\caption{Number of e-foldings $N_e$ versus inflaton field. Here, the largest possible values of $N_e$ are shown and the fast enhancement at the end is due to the field values approaching towards the maxima of the potential ($\phi_{\rm max} = \pi \, f$). Here $f$ varies from $1$ to 3 in the upward direction.}
\label{ne}
\end{center}
\end{figure}

Because the natural inflation potential has a maximum at $\phi = \pi \, f$, depending on the decay constant, there is an upper
limit on $N_e$ which can be realized for a given $f$. It can be shown that even $f = 1$ can generate around $20$ e-foldings as shown in Fig.~\ref{ne}, while $f=2$ can result in a maximal value of $N_e$ around $80$. However, in order to have $|n_s - 1| < 0.05$ one needs larger decay constant.

\subsubsection*{The Spectral Index $n_s$ and Tensor-to-Scalar ratio $r$}
\begin{figure}[h!]
\begin{center}
\includegraphics[width=10cm]{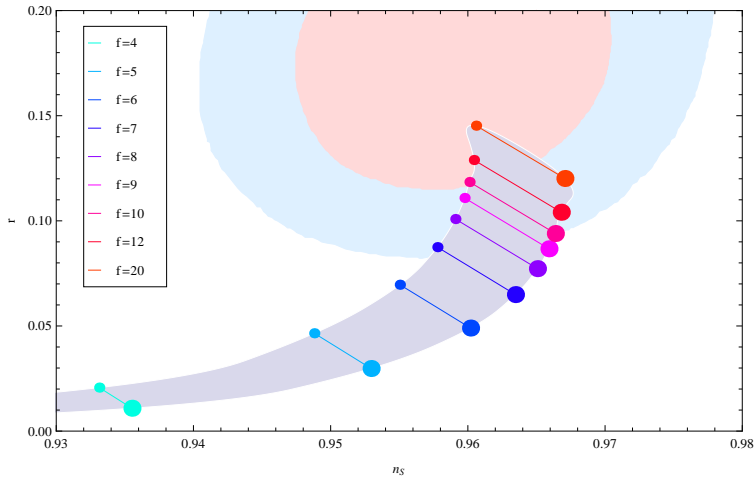}
\caption{  $n_s$ versus $r$  for various effective decay constants from $f=4$ to $f=20$. The blue and red region are respectively the $2 \sigma$ and the
$1\sigma$  regions of $r$ and $n_s$ for BICEP2. The number of  e-folding is from $50$ (small circle) to $60$ (big circle).
}
\label{ns-r}
\end{center}
\end{figure}

Although more than sixty number of e-foldings can be generated even with the decay constant in the range $1< f < 2$, the fitting of the spectral index $n_s$ and tensor-to-scalar ratio $r$ pushes the $f$ window towards $f>4 $. To be more precise, one finds that for $1< f <4$, the spectral index lies in the range $0.1< n_s < 0.9$ while increasing the decay constant values enhances the spectral index. As can be seen from the Fig.~\ref{ns-r}, for e-foldings $50< N_e < 60$, the decay constant $f$ should be in the range of $4 < f < 12 $ in order to be consistent with PLANCK result, and larger than $f > 7 $ in order to fall in the $2 \sigma$ regions of $r$ and $n_s$ for the BICEP2 data.
%and \ref{ns-r-III}, the decay constant should be within $6 < f < 10$ to have the best fit of $n_s$ and $r$.
%\begin{figure}[H]
%\begin{center}
%\includegraphics[width=14.5cm]{r-ns_f-1-5.eps}
%\caption{ $r$ (left) and $n_s$ (right) versus the number of e-foldings $N_e$ for %the  decay constant $f$
% from $1$ to 5 in the upward direction. The two dashed lines correspond to $r = %0.1$ and $n_s = 0.95$, respectively.}
%\label{ns-r-I}
%\end{center}
%\end{figure}
%\begin{figure}[H]
%\begin{center}
%\includegraphics[width=15cm]{r-ns-f_6-10.eps}
%\caption{Plots of $n_s$ and $r$ with a change in the number of efoldings $N_e$ and the decay constant $f$. Here $f$ varies from $6$ to 10 in the upward direction.}
%\label{ns-r-II}
%\end{center}
%\end{figure}
%\begin{figure}[H]
%\begin{center}
%\includegraphics[width=14.5cm]{r-ns.eps}
%\caption{ $r$ (left) and $n_s$ (right) versus the number of e-foldings $N_e$ for %the  decay constant $f$
% from 6 to 10 in the upward direction. }
%\label{ns-r-III}
%\end{center}
%\end{figure}

\begin{figure}
\begin{center}
\includegraphics[width=10cm]{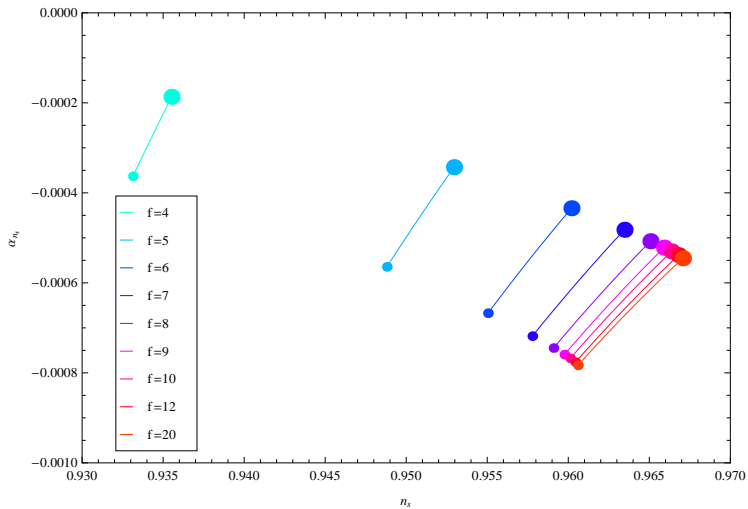}
\caption{  $\alpha_{n_s}$ versus $r$ for decay constants varying from $4$ to $20$.
The number of e-folding is from $50$ (small circle) to $60$ (big circle).}
\label{alphans}
\end{center}
\end{figure}

The running of spectral index $\alpha_{n_s}$ is small. It needs to be of the same order $(10^{-2})$ to reconcile the PLANCK and BICEP2 data eq.(\ref{eq:cosmo-II}). This confrontation has been investigated recently in  \cite{Gong:2014cqa} as can be also seen from Fig. \ref{alphans}. Again, it shows that in order to be consistent with both the PLANCK and BICEP2 data, one needs larger decay constant. However, getting larger decay constant always results in a larger rank of gauge group for gaugino condensation. Of course, $f$ should not be too large. In our case, we constrain the decay constant $f$ to be less than $20$ for a natural choice of parameters.

\subsection*{Benchmark Points in our Aligned Natural Inflation Models}

As shown in Fig.~\ref{ns-r} and the analysis done in the previous subsection, the best fit requirement
for  $n_s$ and $r$ values (from the PLANCK and BICEP2 observations)
demands the decay constant to be within $4 < f_{\rm eff} < 20$. Further, as we have already matched our aligned natural inflationary potential (\ref{eq:Vaxion4}) with the standard form given in (\ref{eq:Vaxion5}), and having all the cosmological observables related analysis been revisited already, now all we need to do is to realize a large decay constant.  For our samplings we would focus in the range $7 < f_{\rm eff} < 12$.

Before  the explicit numerical analysis and sampling of model dependent parameters, let us make the following important points
\begin{itemize}
\item{Although the form of scalar potential suggests that the two fields $\phi_1$ and $\phi_2$ in eq.(\ref{eq:Vaxion2}) or $\psi_1$ and $\psi_2$ in eq.(\ref{eq:Vaxion3}) should be arbitrarily interchangable,  the canonical normalizations fix the choice for a given sampling. This argument is in the sense that the decay constants of the two axions are different
\bea
& & f_1 \equiv f_{c_0} \simeq \frac{g_s}{\sqrt2} \,, \, \, \, \,  f_2 \equiv f_{c^1} \simeq \frac{\sqrt{- 3\, g_s \, \kappa_{B11}} \, \, {\xi_B}^{1/3}}{{\cal V}^{1/3}}~.~\, \nonumber
\eea
In the large volume limit, one naturally expects $f_1 > f_2$, and this hierarchy restricts the interchangeability of the two fields in eq.(\ref{eq:Vaxion2}). The reason for considering this constraint $f_1 >f_2$ is to put a lower bound on the volume of the CY such that ${\cal V} > g_s^{-3/2}$ for natural orientifold constructions. The positive definiteness of kinetic sector demands $\kappa_{B11} < 0$, and some explicit examples of swiss-cheese Calabi-Yau orientifolds with such intersections can be found in \cite{Gao:2013pra}. %For orientifold examples with $\kappa_{B11} =0$, the leading order contribution to the decay constant $f_{c^1}$ scales as ${\cal V}^{-1/2}$ in the CY volume, and so in such cases the canonical normalization of $C_2$ axion changes
. }
%\item{To enjoy the useful features of large volume scenarios, we consider volume of the Calabi-Yau to get stabilized at large enough values ${\cal V} \sim {\cal O}\left(10^2-10^3\right)$.  The relevant model dependent parameters can be consistently chosen. }
%\item We consider the string coupling $g_s$ in the range $0.3\le  g_s \le 1$ to have the decay constant $f_1$ in the range $0.2\le f_1 \le 0.7$.  Further in our sampling we would set $f_2 = 0.01$ without loss of generality.
%The reason for such choice is to keep a mass hierarchy between $c^1$ field and the volume modes $\cV$. This constrain can be relaxed further if we consider the poly-instanton effects~\cite{Gao:2013rra}.
\item As  discussed below eq.(\ref{eq:Vexpand}), if we want to neglected the subleading corrections which are doubly suppressed in $e^{-a_s \, h_1/g_s}$ or $e^{-b_s \, h_3/g_s}$ (or a product of the two factors) with an inherent assumption that $e^{-a_s \, h_1/g_s}\simeq e^{-b_s \, h_3/g_s}$, we have to choose $a_s \, h_1 \simeq b_s \, h_3$ for consistancy, or equivalently $$\frac{h_1}{N_1} \simeq \frac{h_3}{N_2},$$  where $N_1$ and $N_2$ are the ranks of the gauge groups corresponding to the gaugino condensations.
\item Further, while choosing the flux parameters, one has to take care of the requirement of significant suppressions from factors $e^{-a_s \, h_1/g_s}$ as well as $e^{-b_s \, h_3/g_s}$ to trust the hierarchy of masses used for reaching the single field potential. This requirement usually results in a larger value of $h_1$ and $h_3$. %\footnote{If we relax the constrains that $h_1>h_2$ and $h_3>h_4$, we can get desired $eq:samp1eq:samp1f_{eff}$ with rank of gauge group smaller than $32$, however this choice would be rather less natural as $h_1, h_3$ are quadratic while $h_2, h_4$ are linear in gauge fluxes.}.
Also, $h_1$ and $h_3$ should be larger than $h_2$ and $h_4$ from the different flux dependence on $S$ and $G^1$.
\end{itemize}
Several  benchmark points in Table~\ref{table1} and  \ref{table2} have been presented for various model dependent parameters to realize consistent $r$ and $n_s$ values as shown in Fig.~\ref{ns-r}.
%as to make decay constant larger than the ranks of gauge groups. In our sampling we want to demand the rank to be as small as possible. The additional parametric settings are $\Lambda_1 = \Lambda_0,  \Lambda_2 = \delta \, \Lambda_0$ with $\delta = 1/5$. Here, the scalings of $\Lambda_0$ have to be fixed via inflationary Hubble scale $H_{\rm int}$ given in (\ref{eq:cosmo-I}), or equivalently via fixing the scalar power spectrum $P_s$ as per the bound given in (\ref{eq:cosmo-II}). From the BICEP2, the Hubble scale is set to be $H_{\rm inf} \sim 10^{-4} \, M_{Pl}$ and hence in the slow-roll regime one has $V \simeq 3\, H^2 \sim 10^{-8}$.
%Using (\ref{eq:cosmo-III}), this equivalently translates into
%\bea
%P_s \sim \frac{V}{24\, \pi^2 \, \epsilon} \sim 10^{-9}
%\eea
%which with $\epsilon \sim 10^{-2}$ for having large enough $r$ again recovers $V\simeq 10^{-8}$.
%{\bf "Using (\ref{eq:lambda12}), one can consistently fix $\Lambda_0 \sim 10^{-8}$ for overall volume of the Calabi-Yau in the range ${\cal V} \sim {\cal O}(10^2-10^3)$ via  appropriate choice of other parameters", should be more precise if possible.} .
\begin{table}[H]
  \centering
  \begin{tabular}{|c|c|c|c|c|c||c|c|c||c|c|}
  \hline
   & & & & & & & & & & \\
   & $W_{cs}$ & $N_0$ & $A_0$ & $C_{\alpha^\prime}$ & $g_s$& $\ov{\cal V}$ &  $\ov\tau_s$ &  $|V_{\rm LVS}|$& $f_1$ & $f_2$  \\
& & & & & & & & & & \\
\hline
\hline
%& & & & & & & & & & \\
 $S1$ & -12 & 15 & 0.1 & 4.6 & 0.35  & 925.7 & 2.99 & $5.0 \times 10^{-7}$ & 0.248 & 0.105 \\
& & & & & & & & & & \\
 $S2$ & -10 & 3 & 0.4 & 5.1 & 0.35  & 849.6 & 1.68 & $1.8 \times 10^{-7}$ & 0.248 & 0.108 \\
& & & & & & & & & & \\
 $S3$ & -14 & 6 & 0.8 & 7.5 & 0.28  & 421.3 & 2.35 & $6.0 \times 10^{-6}$ & 0.198 & 0.122 \\
& & & & & & & & & & \\
 $S4$ & -20 & 14 & 0.1 & 2.8 & 0.40  & 909.5 & 2.61 & $9.5 \times 10^{-7}$ & 0.283 & 0.113 \\
& & & & & & & & & & \\
 $S5$ & -18 & 16 & 0.1 & 3.5 & 0.30  & 1024.7 & 2.99 & $6.7 \times 10^{-7}$ & 0.212 & 0.094 \\
 & & & & & & & & & & \\
 $S6$ & -11 & 8 & 0.2 & 5.8 & 0.29  & 688.6 & 2.28 & $9.0 \times 10^{-7}$ & 0.205 & 0.105 \\
    \hline
     \end{tabular}
     \caption{The six benchmark samplings for model dependent parameters to stabilize the moduli at large volume minima. Here, $\xi_B = 1 = \xi_S$ and $\kappa_{B11} = -1$ have been used.}
      \label{table1}
\end{table}

\begin{table}
  \centering
  \begin{tabular}{|c|c|c|c|c||c|c||c|c|c|}
  \hline
   & & & & & & & & & \\
   & $A_s$ & $B_s$ & $h_1$ & $h_3$ & $\Lambda_1$& $\Lambda_2$ &  $f_1^\prime$ &  $f_2^\prime$& $f_{\rm eff}$   \\
& & & & & & & & & \\
\hline
\hline
%& & & & & & & & &  \\
 $S1$ & 10 & 1 & 15 & 16 & $1.8 \times 10^{-8}$  & $1.7 \times 10^{-9}$ & 0.078 & 0.078 & 8.131 \\
& & & & & & & & &  \\
 $S2$ & 14 & 3 & 17 & 18 & $1.7 \times 10^{-9}$  & $3.7 \times 10^{-10}$ & 0.069 & 0.069 & 9.452 \\
& & & & & & & & & \\
 $S3$ & 12 &4 & 15 & 16 & $5.2 \times 10^{-9}$  & $1.7 \times 10^{-9}$ & 0.063 & 0.063 & 9.394 \\
& & & & & & & & & \\
 $S4$ & 8 & 1 & 19 & 20 & $9.5 \times 10^{-9}$  & $1.3 \times 10^{-9}$ & 0.071 & 0.072 & 11.035 \\
& & & & & & & & & \\
 $S5$ & 25 & 5 & 18 & 19 & $1.6 \times 10^{-9}$  & $3.6 \times 10^{-10}$ & 0.056 & 0.057 & 8.694 \\
 & & & & & & & & & \\
 $S6$ & 10 & 2 & 13 & 14 & $1.1 \times 10^{-8}$  & $2.0 \times 10^{-9}$ & 0.075 & 0.074 & 7.071 \\
    \hline
     \end{tabular}
     \caption{The manifestation of effective large decay constant and the hierarchial scales $\Lambda_i$'s for the six benchmark samplings presented in Table \ref{table1}. Here, the ranks of  gauge groups are chosen to be $N_1 = 30$ and $N_2 = 32$ while the additional flux parameters are set to be $h_2 = 1 = h_4$.}
      \label{table2}
\end{table}

\section{Open Challenges and Conclusion}
\label{sec_conclusions}

In this paper, we have successfully embedded the idea of KNP \cite{Kim:2004rp} for the enhancement of axion decay constant relevant for realizing the Natural inflation. The inflaton is identified with a linear combination of the universal axion $c_0$ and an involutively odd axion $c^1$. The expressions of decay constants for these two axions enjoy appearance of string coupling $g_s$ and the Calabi-Yau volume ${\cal V}$ with a less suppressed factor as compared to the $C_4$ axions. Moreover, their decouplings in the kinetic sector via the K\"aehler potential are more natural in large volume limit as compared to the case of considering two $C_2$ axionic setup as then, one has to diagonalize the intersection matrix $\kappa_{Sab}$ along the odd directions $a$ and $b$. Despite of the several nice features of our model, there are certain assumptions to be consistently realized in concrete setups, especially on the technical grounds. On these lines, let us recall that the original universal axion monodromy inflation \cite{Blumenhagen:2014gta} has two delicate issues as below
\begin{itemize}
\item{The decay constant for universal axion $c_0$ is given as $f_{c_0} = \frac{g_s}{\sqrt2}$, and so natural inflation embedding demands string coupling to be in the window $1< g_s < 10$ and thus pushing the whole description into the non-perturbative regime.
    %where infinitely many (un-)known corrections have to be taken care of. There has been argued to still have the possibilities of having control over the whole framework in the large volume limit given that those corrections appear with Calabi-Yau volume suppression factors.
    Our approach of realizing the KNP-type inflation with inflaton being a combination of the universal axion $c_0$ and the odd axion $c^1$ provides a natural way of enhancing the decay constant in the regime where the perturbative description remains trustfully valid along with the support of large volume scenarios.}
\item{The second delicate assumption of inflationary model in \cite{Blumenhagen:2014gta} is related to facilitate a hierarchy in the dilaton and universal axion at the
 tree-level superpotential. This flux superpotential depends on the landscape of background fluxes and it would be interesting to construct the explicit models in which this requirement could be satisfied.}
\end{itemize}
In addition to the second point, it would be interesting to address more technical issues like the tadpole/anomaly cancellations in  concrete Calabi-Yau orientifold examples with all the suitable gauge fluxes arranged through the incorporation of relevant involutively odd two-cycles  to  contribute the non-perturbative effects. Further, the trans-Planckian nature of the inflaton opens up some more challenges and hence there are some cautionary concerns on the lines of \cite{Chialva:2014rla, Baumann:2011nm, Burgess:2014tja, Martin:2000xs, Brandenberger:2004kx, Brandenberger:2012aj}. One of such concerns could be the inflaton coupling to the gauge degrees of freedom living on the two stacks of $D7$-brane wrappings with magnetic-fluxes turned-on, and those could be of the following kind
\bea
& &\hskip-2cm  {\cal L} \supset \frac{\phi_1}{f_1} \, \biggl[\frac{a_s \, h_1}{32 \, \pi^2} \, F_{\mu \nu 1} \, F^{\mu \nu 1}+\frac{b_s \, h_3}{32 \, \pi^2} \, F_{\mu \nu 2} \, F^{\mu \nu 2} \biggr] \\
& &+\frac{\phi_2}{f_2}  \biggl[\frac{a_s \, h_2}{32 \, \pi^2} \, F_{\mu \nu 1} \, F^{\mu \nu 1}+\frac{b_s \, h_4}{32 \, \pi^2} \, F_{\mu \nu 2} \, F^{\mu \nu 2} \biggr] ~,~\nonumber
\eea
where $a_s = \frac{2\pi}{N_1}$ and $b_s = \frac{2\pi}{N_2}$ with $N_i$'s being the ranks of gauge groups. In order to avoid the gauge degrees of freedom being supermassive by acquiring the trans-Planckian masses out of the axion vacuum expectation values, one has to ensure that the overall coupling still remains under
control. For that one has to have large rank of gauge groups which comes out to be an unnatural requirement beyond a certain value. Also in our setup, we have required some of gauge flux parameters ($h_1$ and $h_3$) to be relatively large (of order 10) to sustain the mass-hierarchy, and this has to be done in a consistent manner by not letting these fluxes become very large.%, via keeping $a_s h_1$ or $b_s h_3$ less than unity.  Nevertheless, the whole setup should not be overloaded by the fluxes to destroy the moduli stabilization. Although we have some improvement in the sense that we do not take the ranks of gauge groups to be very large, they still are of the order 100 if we want string coupling around $g_s \le 0.2$.
Although in our samplings we have successfully realized large effective decay constant with the rank of the gauge groups being not too large ($N_1 = 30$ and $N_2 = 32$),  it would be interesting to increase the number of the odd axions and take the requirement for the rank of gauge group to be below ten on the lines of \cite{Choi:2014rja}. It suggests an exponential enhancement of the decay constant with increasing the number of axions
 in the KNP formalism.

\section*{Acknowledgments}

We would like to thank Ralph Blumenhagen and Anupam Mazumdar very much for helpful discussions. We also thank Michele Cicoli and Erik Plauschinn for useful discussion. TL and XG were supported in part by the Natural Science Foundation of China under grant numbers 10821504, 11075194, 11135003, and 11275246, and by the National Basic Research Program of China (973 Program) under grant number 2010CB833000. PS was supported by the Compagnia di San Paolo contract ``Modern Application of String Theory'' (MAST) TO-Call3-2012-0088.

%\clearpage
\nocite{*}
\bibliography{Axion}
\bibliographystyle{utphys}

\end{document}